\documentclass{article}

\usepackage[T1]{fontenc}
\usepackage{float}
\usepackage{svg}
\svgpath{{../Figures/}}
\usepackage{amsmath}
\usepackage{url}
\usepackage{graphicx} \usepackage{float} \usepackage{subfigure} \usepackage{booktabs} \usepackage{array}
\usepackage{tabu}
\usepackage{booktabs}
\usepackage{amsmath} \usepackage{amsopn} \usepackage{amssymb} 
\usepackage{chemformula}
\usepackage{times}
\usepackage{dblfloatfix}
\usepackage{arydshln}
\usepackage[pagestyles,raggedright]{titlesec}
\usepackage{indentfirst} 
\usepackage{geometry}
\usepackage{gensymb}
\usepackage[breaklinks=true, pdfpagemode=UseOutlines, colorlinks=true, linkcolor=black, filecolor=black, urlcolor=black, citecolor=black, plainpages=false, hypertexnames=false, pdfpagelabels=true, hyperindex=true]{hyperref}
\usepackage[numbers,sort&compress]{natbib}
\usepackage{wrapfig}

\usepackage[super]{nth}

\usepackage{multicol,lipsum}
\usepackage{color,soul}

\usepackage{eqnarray,amsmath}
\definecolor{HeaderLine}{RGB}{0, 131, 162}
\newpagestyle{main}{
   
}
\DeclareMathOperator*{\argmin}{arg\,min}

\titleformat{\section}{\normalfont\fontsize{10}{10}\bfseries}{\thesection}{1em}{}

\begin{document}
% An Interpretable Boosting-based Predictive Model for Estimating Transformation Temperatures of Shape Memory Alloys with Feature Engineering

\title{\LARGE{\textbf{An Interpretable Boosting-based Predictive Model for Transformation Temperatures of Shape Memory Alloys}}}
\author{\textbf{Sina Hossein Zadeh}$^{1*+}$, \textbf{Amir Behbahanian}$^{1*}$, \textbf{John Broucek}$^1$, \\
\textbf{Mingzhou Fan}$^2$, \textbf{Guillermo Vazquez Tovar}$^1$, \textbf{Mohammad Noroozi}$^3$,
\textbf{William Trehern}$^1$,\\
\textbf{Xiaoning Qian}$^2$, \textbf{Ibrahim Karaman}$^1$, \textbf{Raymundo Arroyave}$^1$\\
$^1$ Department of Materials Science \& Engineering, Texas A\&M University \\
$^2$ Department of Electrical \& Computer Engineering, Texas A\&M University \\
$^3$ Department of Industrial \& Management, University of South Florida \\
$^*$ These authors contributed equally.
\\
$^+$ E-mail address of the corresponding author: sina@tamu.edu Phase Transformation, Feature Engineering}
\date{}
%=============================================
\maketitle
\thispagestyle{main}
%=============================================
% please write your abstract

\begin{abstract}
In this study, we demonstrate how the incorporation of appropriate feature engineering together with the selection of a Machine Learning (ML) algorithm that best suits the available dataset, leads to the development of a predictive model for transformation temperatures that can be applied to a wide range of shape memory alloys. We develop a gradient boosting ML surrogate model capable of predicting Martensite Start, Martensite Finish, Austenite Start, and Austenite Finish transformation temperatures with an average accuracy of more than 95\% by explicitly taking care of potential distribution changes when modeling different alloy systems. We included heat treatment, rolling, extrusion processing parameters, and alloy system categorical features in the model input features to achieve more accurate and realistic results. In addition, using Shapley values, which are calculated based on the average marginal contribution of features to all possible coalitions, this study was able to gain insights into the governing features and their effect on predicted transformation temperatures, providing a unique opportunity to examine the critical parameters and features in martensite transformation temperatures.

\end{abstract}

\begin{multicols}{2}

\section{Introduction}

Shape memory alloys (SMAs) exhibit reversible martensitic transformations from a low-temperature, low-symmetry martensite phase to a high-temperature, high-symmetry austenite phase \cite{ma2010high}. As a result of this reversible transition, they exhibit several behaviors, including the Shape Memory Effect (SME) \cite{1975ShapeAlloys, Otsuka1986PseudoelasticityAlloys}, superelasticity \cite{ Otsuka1986PseudoelasticityAlloys, ma2010high}, elasto-/magneto-caloric effects \cite{snodgrass2019multistage, manosa2010giant,bechtold2012high,tuvsek2015elastocaloric,tuvsek2016understanding,kirsch2018niti,sehitoglu2018elastocaloric,liu2012giant}, tunable thermal expansion \cite{gehring2020evolution,monroe2016tailored}, stiffness \cite{ma2017self,ma2020variably}, magnetic exchange bias \cite{sharma2020martensitic}, and spin glass transitions \cite{binder1986spin,morito1998magnetic,chatterjee2009reentrant}.
\par
Advances in SMAs are intermittent. A significant discovery has only occurred once a decade since the late 1960s. For example, the discovery of NiTi in the 60s \cite{buehler1963effect, Wang1965CrystalTiNi}, Cu-based SMAs in the 1970s \cite{Wayman1980SomeAlloys}, superelasticity in the 1980s \cite{Korotaev1984SuperelasticityTwinning, Yoneyama1989EvaluationWire.}, magnetic SMAs in the 1990s \cite{Wuttig2000Occurrenceinvited}, $\beta$-Ti SMAs in the 2000s \cite{Inamura2005MechanicalGa, Masumoto2006EffectsAlloys, PING2006TEMAlloy, canadinc2019ultra, Wang2008TheAlloy, Wang2010InAlloys}, and multi-component high-temperature SMAs in the 2010s \cite{canadinc2019ultra}. There is a need to develop a methodology that will accelerate the discovery of new SMAs, thereby reducing the discovery cycle from decades to just a few months.   
\par
Conventional alloy discovery has relied on a tradeoff system where the gain in one target leads to another worsening. Due to the numerous tradeoffs that often must be employed in such systems, optimizing multiple characteristics becomes even more challenging. It primarily targets binary systems where interactions between elements are often known or at least predictable. In conventional alloys, the quest for finding tradeoffs to tune specific characteristics has motivated the exploration of the multi-principal element, or high entropy alloy design space \cite{yeh2004nanostructured,tsai2014high,zhang2014microstructures,miracle2017critical, Zadeh2020RegulatingDeformation}. A similar paradigm has emerged in functional alloys with several compositionally complex SMAs, and Multi-Principal Element Multi-Functional Alloys (MPEMFA), which may demonstrate superior performance compared to less complex counterparts \cite{firstov2015high}. As an example of these quests, recently discovered quaternary and quinary MPEMFAs exhibit reversible thermally induced martensite transformation, SME, and superelasticity above 700°C, the highest temperature superelasticity reported to date \cite{wang2019superelastic}.

Efficient discovery of new compositionally complex alloys is challenging due to the vastness of the chemical and microstructural space and extreme sensitivity to minute changes in chemistry and/or microstructure \cite{umale2019effects}. The successful discovery of MPEMFAs is at the expense of exploring a very high-dimensional, non-linear, and highly singular mathematical space, in which querying a single point experimentally comes at a high cost.

Expanding the chemical discovery space may lead to new discoveries, but it may also result in undesirable properties. This is especially true when considering their solid-state actuation properties and martensitic transformation temperature characteristics for optimizing actuation properties under transformation temperature constraints. As an example, in TiPt and Pd systems, higher alloy complexity led to smaller irrecoverable/permanent strains and actuation thermal cycling, but at the expense of a larger thermal hysteresis \cite{Yamabe-Mitarai2020TiPd-Advances}. As chemical complexity increases, atomic interactions make the understanding of the new design space more complicated as numerous factors may be able to yield the same result. 

Current alloy development frameworks are insufficient, as simulation-assisted frameworks can be computationally expensive, do not readily incorporate data from experiments, and are sequential. Traditional high throughput (HTP) combinatorial approaches are incapable of dealing with the high dimensionality and complexity of the MPEFMA space. In addition, HTP approaches are ‘one-shot’ or ‘open loop’ schemes without an iterative framework, and are suboptimal in resource allocation. State-of-the-art machine learning (ML)/AI-driven approaches may be unsuitable in cases where the materials space is highly non-linear, sparsely queried, or very singular. The use of modeling, combined with high throughput experiments, has the potential to vastly increase the rate of discovery in these new systems due to the newfound ability to apply active learning methods to find trends in small elemental variations, which could never be observed before.

Avoiding computationally expensive calculations such as first-principles calculations \cite{chakraborty2016unraveling, wang2018martensitic}, or theoretical studies \cite{platl2020determination, dai2004design, Narayana2018EstimationAlloys}, multiple research groups have tried to use classical machine learning~(ML) models and, more recently deep neural networks to create surrogate models, connecting chemistry to transformation temperatures. Most of the models are fitted to small training data sets that are less diverse in terms of the included elements and different SMA systems, limiting the application space of these models \cite{mehrpouya2021prediction, xue2017informatics, zhang2020transformation, chen2022thermodynamic, huang2020combined, Narayana2018EstimationAlloys, Tian2022Machine-learningAlloys}. On the other hand, some data sets are enriched with computational data stemming from simulation models such as CALPHAD \cite{sundman2007computational}, which limit the results to the uncertainties and assumptions of the underlying calculations \cite{mu2021predicting, peng2020coupling}. 

Although the importance of processing parameters on martensitic transformation is known, only limited studies \cite{zhang2020transformation, xue2017informatics, liu2021physics, mehrpouya2021prediction} have considered the effect of processing parameters in their model to avoid the complexities added by these features. Wang \emph{et. al.} in their recent work~\cite{Wang2021CompositionallyPredictions} attempted to cover the shortcomings of the literature and have combined composition-based featurization with a fractional encoding of the materials. However, the complexity of their model makes tracking and providing feature importance calculations a cumbersome task. In addition, their model limits the feature space to chemistry-related information, and processing parameters are not included. As a result, having a model that covers several alloy systems and is based on a relatively diverse data set, which considers the features containing the composition-based information and the manufacturing parameters, is missing in the literature. 

Considering the aforementioned missing pieces to have a complete and comprehensive model, we are structuring our work to build a surrogate model that can fit the multi-dimensionality of a diverse yet limited data set.  The precise feature engineering enabled the possibility of making predictions for various alloy systems. In addition, the generated features and Shapley values provided the chance to shed light on the physics of the transformation.

Our work is structured as follows; We first describe the data set, the background of the ML algorithm, and the surrogate ML model in the methods section. Subsequently, we will discuss our results and the novelty of this work providing physics-related information about the features affecting the transformation temperatures.

\begin{figure*}[ht]
  \centering
  \includegraphics[width =\textwidth]{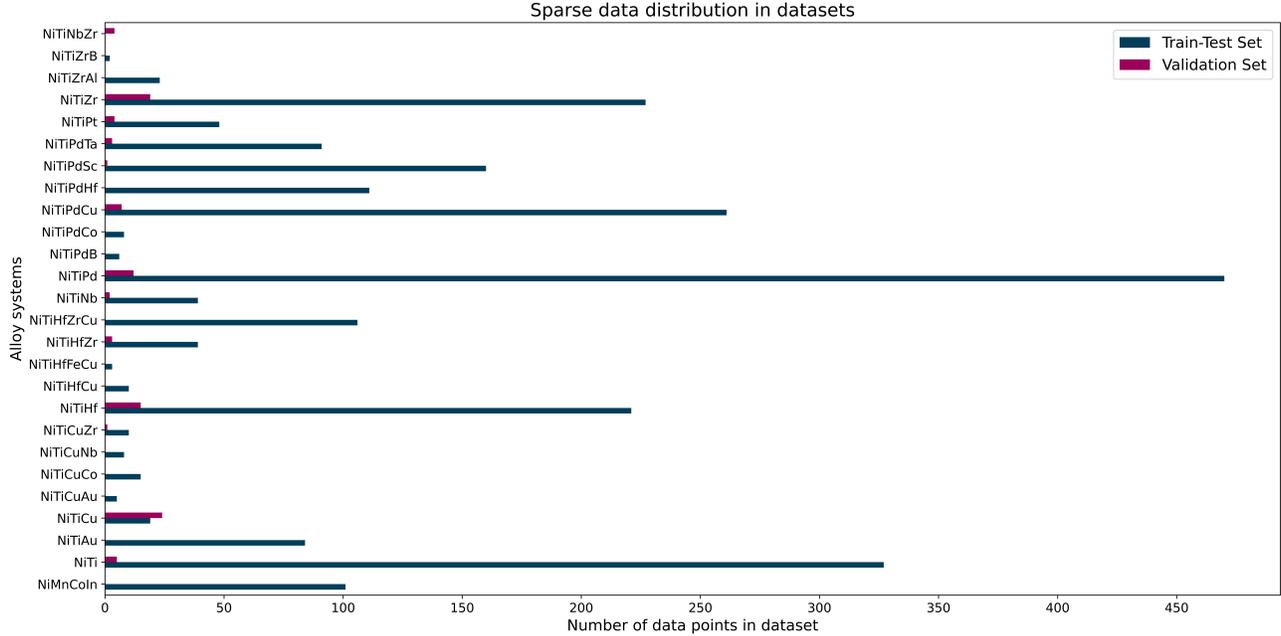}
  \caption{The number of data points available for each alloy system in the database. Please note that each alloy system is composed of multiple compositions, and each composition might have multiple data points with different processing parameters. The figure was plotted using Matplotlib \cite{Hunter2007Matplotlib:Environment}.}
  \label{fig:comp}
\end{figure*}
\section{Methods}

In this section, we first elaborate on the structure of the data set under the study, data pre-processing, and feature generation procedures to make our reported results reproducible. More importantly, we introduce the basics of the adopted ML surrogate model based on CatBoost, for which we will  demonstrate its generalizable prediction performance with respect to the corresponding evaluation metrics in our experiments in Section~\ref{sec:exp}. Last but not least, by adopting the Shapley value~\cite{NIPS2017_8a20a862}, we also provide the feature importance analysis to interpret the learned CatBoost model for SMAs transformation temperatures. 

\subsection{Data Preparation}\label{sec:DP}
A database with information on composition, processing parameters, and thermodynamic responses was constructed by combining the data mined from the literature, and our in-house experimental results \cite{trehern2022data}. Composition data was represented in atomic percentages at the resolution of 0.1at\%. Processing parameters were used, including the parameters in heat treatment, rolling, and extrusion. Both composition and processing parameters construct an initial feature list for the ML surrogate model. Thermodynamic responses were limited to transformation temperatures as these are the targets for this study. 

After cleaning the data (i.e., removing data points with duplicated and NaN values) and removing inconsistent values (i.e., removing data points where atomic percentages exceeded 100, Martensite Start (Ms) $>$ Austenite Finish (Af), Martensite Finish (Mf) $>$ Austenite Start (As), As $>$ Af, and Mf $>$ Ms), 2,494 data points for 237 different shape memory alloy were acquired. These data points have 13 different processing parameters. The dataset in the compositional space has data concentration around specific compositions. For some compositions, we had as low as one experimental measurement data point per composition; for others, there were more than ten data points per composition. This fact creates an under-sampling of the composition space for some regions. Still, it doesn't imply an imbalance in the dataset, as the data points with the same composition have different processing parameters. To add another layer of model evaluation for generalizability, we divided the data set into the train/test set and the validations set. The train/test set contains compositions with at least two experiments per composition and is divided into training and test sets for appropriate training of the model and to avoid overfitting the training data. On the other hand, the validation set is the set of compositions with only one experiment. The set is used to evaluate the model further. The number of samples available for alloy systems in each data set is shown in Fig.\ref{fig:comp}. \par

We employed three featurization methods for the present work: processing-related, composition-related, and alloy-system-related. Compositions were featurized with the Jarvis database \cite{Choudhary2020TheDesign} using the \texttt{CBFV} \cite{Kauwe2018MachineInorganics, Kauwe2020CanMaterials, Wang2020MachinePractices, Clement2020BenchmarkLearning, Murdock2020IsProperties, Wang2021CompositionallyPredictions} Python \cite{10.5555/1593511} library. We added these features based on previous experiments and density functional theory, ML, and classical force-field calculations, as a result, increasing the total number of features to 3,079. For more detailed explanations of the features generated in this paper, please refer to JARVIS-Tools \cite{JARVIS-ToolsDocumentation} and CBFV \cite{Kaaiian/CBFV:Vector} documentation. \par 
We note that our final feature set also contains eight categorical features to help our ML surrogate model better predict thermodynamic properties for different SMA systems. These categorical features are NiTi-based, NiTiPd-based, NiTiCu-based, NiTiHf-based, NiTiNb-based, NiTiZr-based, NiTiB-based, and alloy systems. This set of categorical features provides information on SMA systems with their main element combinations. As an example, the alloy system for Cu2Ni3Pd5Ti10 is NiTiPdCu. Other categorical features are subcategories of alloy systems. In the case of Cu2Ni3Pd5Ti10, it belongs to the NiTi-based, NiTiPd-based, and NiTiCu-based categories, so the value for these features is \textit{True}, and the value for the four remaining categorical features (i.e. NiTiHf, NiTiNb, NiTiZr, and NiTiB-based) is \textit{False}. As a result of feature engineering, the database contained 3,087 features ready for training. 

\subsection{CatBoost Surrogate Model}
The number of alloy systems, significant change in thermodynamic targets with slight composition variation, number of features, and presence of categorical and numerical features in the database indicate a complex non-linear learning space. Decision-tree-based ML models are good candidates for capturing nonlinear composition\textbackslash processing\textbackslash structure property-space mapping relationships in a high-dimensional learning space. Different boosting and bagging strategies, built upon tree-based weak learners, such as Random Forests \cite{Breiman2001RandomForests}, XGBoost \cite{Chen2016XGBoost:System}, LightGBM \cite{NIPS2017_6449f44a}, and CatBoost \cite{Prokhorenkova2018CatBoost:Features}, have been developed to achieve state-of-the-art prediction performance. Convenient in working with categorical data and having multiple targets are the other reasons that we chose CatBoost as our final ML surrogate model. In the following subsections, we introduce the background of the ML algorithm, model evaluation methods, and hyperparameter tuning by Bayesian optimization~(BO).

\subsubsection{Background of the ML Algorithm}
Boosting~\cite{Freund1999ABoosting} is a family of ML algorithms that combines multiple weak learners to build a strong learner for prediction. Decision trees~\cite{Song2015DecisionPrediction.}, which recursively partition the input space and learn the decision rules for each subspace based on training data, are usually chosen as weak learners or base learners because of their lightweight nature and low computational costs.

Gradient boosting \cite{Bentejac2021AAlgorithms}, as a member of the boosting family, utilizes the gradient of a pre-defined loss function to combine the weak learners.
Unlike other implementations of gradient boosting, including  XGBoost~\cite{Chen2016XGBoost:System} and LightGBM~\cite{NIPS2017_6449f44a}, CatBoost~\cite{Prokhorenkova2018CatBoost:Features} tries to fight potential prediction shifts with an ordered boosting method and has been reported to achieve more reliable prediction performance.

In CatBoost, the strong learner $M_{t}$ is updated as~\cite{Prokhorenkova2018CatBoost:Features}:
$$M_{t+1}(\textbf{x}) = M_t(\textbf{x}) - \alpha h^*_{t}(\textbf{x}),$$
where $h^*_{t}(\textbf{x}) \in \argmin_{h(\textbf{x})}(E_\textbf{x}[(h(\textbf{x}) - \frac{\partial L(f, \textbf{x}, y)}{\partial f}|_{f = M_t})^2])$ is the estimation of $\frac{\partial L(f, \textbf{x}, y)}{\partial f}|_{f = M_t}$ with the weak learner and $\frac{\partial L(f, \textbf{x}, y)}{\partial f}|_{f = M_t}$ is the gradient of the loss function w.r.t the current strong learner $M_t$,

$\alpha$ is the learning rate to control the update step size. $L(f, \textbf{x}, y) = \mathcal{L}(f, \textbf{x}, y) + J(f)$, which is the loss function connecting the prediction to the target, usually contains an L2 regularization term $J(f)$ to prevent from overfitting.\par
As another way to resist prediction shift and speed up the training procedure, Bayesian bootstrap~\cite{rubin1981bayesian} is applied in CatBoost in each iteration on the training dataset to sub-sample and re-weight the data and balance model inputs. 

\begin{table*}[ht]
  \centering
  
  \begin{tabular}{@{}ccc@{}}
    \hline
    \multicolumn{3}{c}{CBFV Python library} \\
    \toprule
    Parameters  & Values & Comment    \\
    \midrule
    elem\_prop & jarvis & \begin{tabular}[c]{@{}l@{}}
The other good option is 'oliynyk'. It generates fewer features; hence it is \\ computationally cheaper. However, we got slightly better results using 'jarvis'.
\end{tabular} \\
    \addlinespace[1ex]
    \hdashline[0.5pt/5pt]
    \addlinespace[1ex]
    extend\_features & True & -   \\
    \addlinespace[1ex]
    \hdashline[0.5pt/5pt]
    \addlinespace[1ex]
    sum\_feat & True & \begin{tabular}[c]{@{}l@{}}
It generates 438 more features when used with jarvis.
\end{tabular} \\   \\
     \hline
    \multicolumn{3}{c}{CatBoost Python library} \\
    \toprule
    Parameters  & Values & Comment    \\
    \midrule
    boosting\_type & Plain & The classic gradient boosting scheme.  \\
    \addlinespace[1ex]
    \hdashline[0.5pt/5pt]
    \addlinespace[1ex]
    learning\_rate & 0.2092 & Boosting learning rate. \\
    \addlinespace[1ex]
    \hdashline[0.5pt/5pt]
    \addlinespace[1ex]
    depth & 6 & Depth of the tree. \\
    \addlinespace[1ex]
    \hdashline[0.5pt/5pt]
    \addlinespace[1ex]
    l2\_leaf\_reg & 0.05992 & Coefficient at the L2 regularization term of the cost function. \\
    \addlinespace[1ex]
    \hdashline[0.5pt/5pt]
    \addlinespace[1ex]
    bagging\_temperature & 0.5562 & 
    Assigning random weights to objects by using the Bayesian bootstrap. \\
    \addlinespace[1ex]
    \hdashline[0.5pt/5pt]
    \addlinespace[1ex]
    loss\_function & MultiRMSE & The metric to use in training. \\
    \addlinespace[1ex]
    \hdashline[0.5pt/5pt]
    \addlinespace[1ex]
    eval\_metric & MultiRMSE & The metric used for overfitting detection. \\

    \bottomrule
  \end{tabular}
  
  \caption{Software parameters were used to build the ML model.}
  \label{table:1}
\end{table*}
\subsubsection{Loss Function and Evaluation}

The choice of the loss function $\mathcal{L}(f, \textbf{x}, y)$ is critical in the training process to achieve accurate predictions. 
To measure the prediction accuracy for a single target, Root Mean Square Error~(RMSE), Mean Absolute Error (MAE), and coefficient of determination~(R2 score) are broadly used. The R2 score, measuring the dependency relationship between the prediction and target, is defined as \cite{Kvalseth1985Cautionary2}:
$$R^2 = 1 - \frac{\sum_i (y_{i} - M(\textbf{x}_i))^2}{\sum_i (y_{i} - \bar{y})^2},$$
where $\bar{y}$ is the average over all the target values. 

RMSE measures the difference between the target and the prediction by~\cite{Chai2014RootLiterature}:
$$\text{RMSE}(M(\textbf{x}), y) = \sqrt{\frac{\sum_{i}(M(\textbf{x}_i) - y_{i})^2}{N}},$$
where $M(\textbf{x}_i)$ is the prediction of the $i$-th sample, $y_{i}$ is the corresponding target, and $N$ is the number of samples.
MAE, as another broadly used metric, is defined as~\cite{Chai2014RootLiterature}:
$$\text{MAE}(M(\textbf{x}), y) = \frac{\sum_{i}|M(\textbf{x}_i) - y_{i}|}{N}.$$
A perfect predictor (always correct) would have a $0$ RMSE and MAE with an R2 score of $1$. \par
When it comes to multiple target prediction, we can adopt RMSE to MultiRMSE as follows: 
$$\text{MultiRMSE}(M(\textbf{x}), y) = \sqrt{\frac{\sum_{i}\sum_j(M(\textbf{x}_i)_j - y_{i, j})^2}{N}},$$
where $M(\textbf{x}_i)_j$ is the prediction of the $j$-th target of the $i$-th sample ($\textbf{x}_i$), $y_{i, j}$ is the corresponding target, and $N$ is the number of samples. 

\subsubsection{Hyperparameter Tuning}
Fine-tuning the model's hyperparameters can be a pathway to pursuing higher prediction accuracy.  Before training the model, we tuned the hyperparameters of the CatBoost algorithm using the BO approach \cite{Frazier2018AOptimization}, utilizing the \texttt{Bayesian Optimization} \cite{Nogueira2014BayesianPython} Python library. The optimization process was performed on the training set, and the negative square root of the \texttt{mean\_squared\_error} was set as the target of the BO algorithm to optimize. The resulting parameters used to build the ML model are illustrated in Table~\ref{table:1}. \par

By considering the hyperparameters as the input to a black-box function whose output is the corresponding model performance, the task of fine-tuning the hyperparameters can be viewed as a black-box optimization problem. BO~\cite{Frazier2018AOptimization} that iteratively queries new input and eventually optimizes the prediction performance is suitable under this scenario.
BO models the black-box performance function as a Gaussian process~(GP) $\mathcal{GP}(\mu(\cdot), Cov(\cdot, \cdot))$. An acquisition function guides the selection of new query points based on the updated posterior of the GP condition on the current observations.

The choice of acquisition function is critical and affects the optimization efficiency. Expected Improvement~(EI), Entropy Search~(ES), and Upper Confidence Bound~(UCB) are examples of well-studied acquisition functions~\cite{Frazier2018AOptimization}. We focus on the UCB acquisition because of its low computational cost and strong performance, which takes the form of:
$$\alpha(x) = \mu(x|\mathcal{D}) + \beta \sqrt{Cov(x, x|\mathcal{D})},$$ where $\mathcal{D}$ denotes the set of current observations, $\mu(x|\mathcal{D})$ and $Cov(x, x|\mathcal{D})$ represent the mean and variance of the updated posterior GP, respectively. UCB balances the exploration and exploitation by the hyperparameter $\beta$, where higher $\beta$ values lead to a higher level of exploration while lower $\beta$ lead to exploitation more. In our experiments, $\beta=2.576$.

\subsection{SHAP-based Model Interpretation} 

Besides prediction accuracy, model interpretability is another goal when developing AI/ML surrogate models, particularly for scientific machine learning (SciML), as in this paper. To better understand the composition-processing-property relationships, we also calculate the SHapley Additive exPlanation (SHAP)~\cite{NIPS2017_8a20a862} for each feature, enabling information-theory-based feature prioritization. The corresponding SHAP values capture the importance of the related features in determining materials' thermodynamic responses, which provides insight into SMA physics. 

The SHAP Value $\phi(k, M)$ for the feature $k$ in a model $M$ is defined as:
$$\phi(k, M) = \sum_{S \subset F/k} w_{S, F}[M^{S\cup\{k\}}(\textbf{x}^{S\cup\{k\}}) - M^{S}(\textbf{x}^{S})],$$
Where $F$ is the set of all the features, $M^{S}(\textbf{x}^{S})$ is the retrained model with only the features in a subset $S$ instead of all the features.

The retraining of $2^{|F|-1}$ models to calculate the SHAP value of each feature, however, is computationally infeasible in most applications. The original paper~\cite{NIPS2017_8a20a862} proposed several approximation methods to tackle this exponential computational complexity challenge with significantly reduced retraining time. 
SHAP Value can be viewed as a weighted average of the difference between the retrained model with and without feature $k$ for each subset $S$, providing a quantitative feature importance metric capturing its contribution to the final model prediction.

\section{Results and Discussion}
\label{sec:exp}
As detailed in Section \ref{sec:DP}, we have collected 2,494 data points for a set of 237 SMA compositions composed of 26 different alloy systems. The training was performed on 80\% of the training/test data set, {which contains 137 SMAs from NiMnCoIn, NiTi, NiTiAu, NiTiCu, NiTiCuAu, NiTiCuCo, NiTiCuNb, NiTiCuZr, NiTiHf, NiTiHfCu, NiTiHfFeCu, NiTiHfZr, NiTiHfZrCu, NiTiNb, NiTiPd, NiTiPdB, NiTiPdCo, NiTiPdCu, NiTiPdHf, NiTiPdSc, NiTiPdTa, NiTiPt, NiTiZr, NiTiZrAl, and NiTiZrB alloy systems,} and the performance of the model was evaluated on the test set (20\% of the above data set). Furthermore, a ``validation'' set that contains 100 SMAs from NiTi, NiTiCu, NiTiCuZr, NiTiHf, NiTiHfZr, NiTiNb, NiTiPd, NiTiPdCu, NiTiPdSc, NiTiPdTa, NiTiPt, NiTiZr, and NiTiNbZr alloy systems was held out, excluding from the above training/test data set to better evaluate the generalizability of the trained CatBoost surrogate. 
\begin{figure}[H]
  \centering
  \includegraphics[width = 0.5\textwidth]{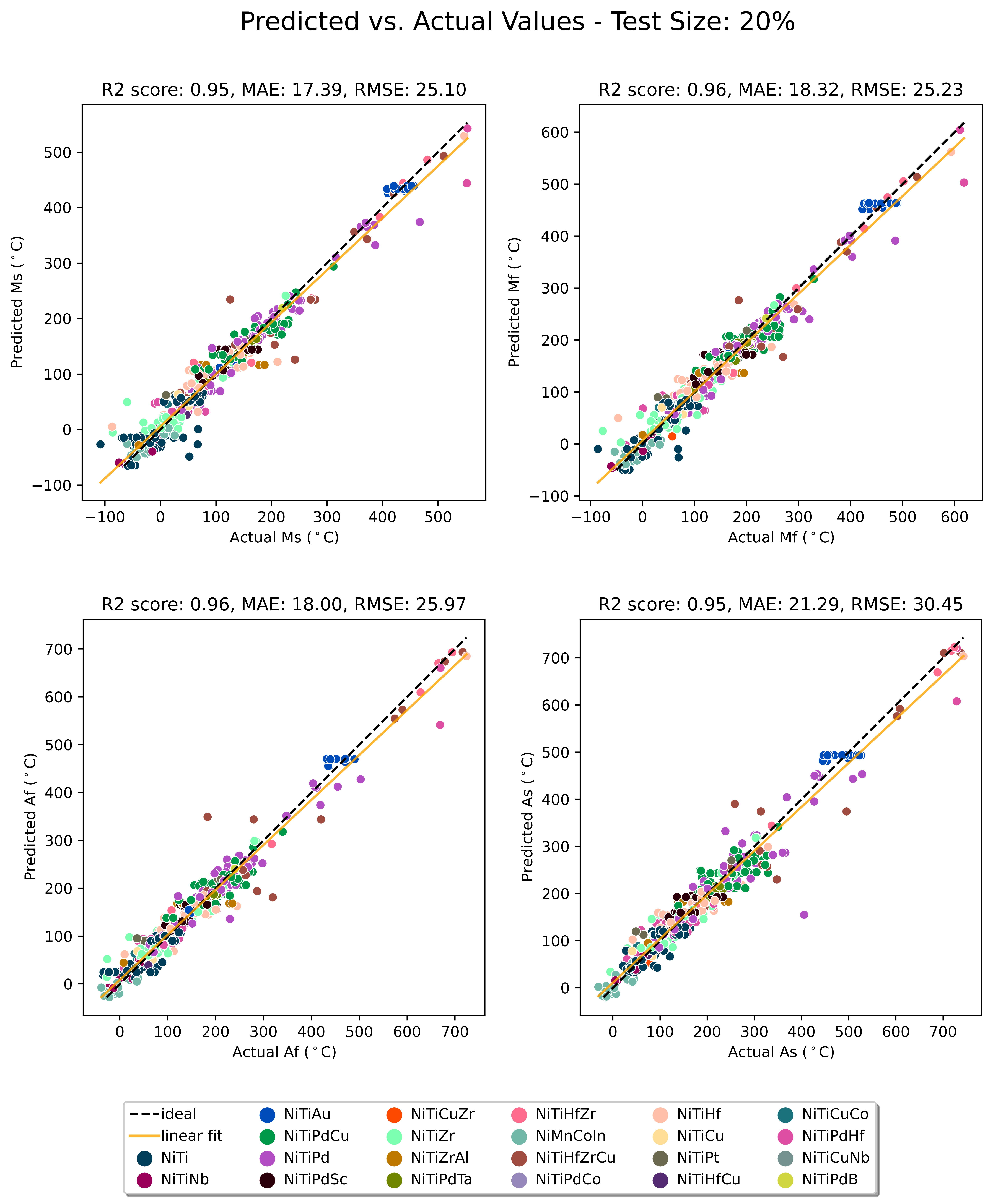}
  \caption{Predicted and actual values of Ms, Mf, As, and Af temperatures for the test set. The figure was plotted using Matplotlib and seaborn \cite{Waskom2021Seaborn:Visualization}. The visually distinct colors generated by iWantHue \cite{IWantHue}.}
  \label{fig:test}
\end{figure}
The average $R^2$ score, MAE, and RMSE of the model predictions were adopted for performance evaluation, calculated using scikit-learn \cite{JMLR:v12:pedregosa11a} \texttt{r2\_score}, \texttt{mean\_absolute\_error}, and \texttt{mean\_squared\_error} functions. On the training set (80\% of the training/test data set), the average $R^2$, MAE, and RMSE values of Ms, Mf, Af, and As temperature predictions are 0.95, 18.47$^{\circ}C$, and 29.97$^{\circ}C$. The same evaluation metrics for the test set (20\% of the training/test data set) are 0.95, 18.75$^{\circ}C$, and 26.78$^{\circ}C$. The high $R^2$ score for the training and test set

is an indication of not having under/overfitting for the trained model. Figure~\ref{fig:test} shows the model's performance for the test set. The $R^2$ scores for Ms, Mf, Af, and As temperatures are 0.95, 0.96, 0.96, and 0.95 respectively. The MAE for the same targets is 17.39$^{\circ}C$, 18.32$^{\circ}C$, 18.00$^{\circ}C$, and 21.29$^{\circ}C$. The RMSE for the same transformation temperatures are 25.10$^{\circ}C$, 25.23$^{\circ}C$, 25.97$^{\circ}C$, and 30.45$^{\circ}C$. Also, when we look into more detailed thermodynamic predictions for different alloy systems, it is clear that our CatBoost model with customized input features not only correctly predicted transformation temperatures of NiTi-based SMAs but also did a good job predicting NiMn-based alloys. \par

\begin{figure}[H]
  \centering
  \includegraphics[width = 0.5\textwidth]{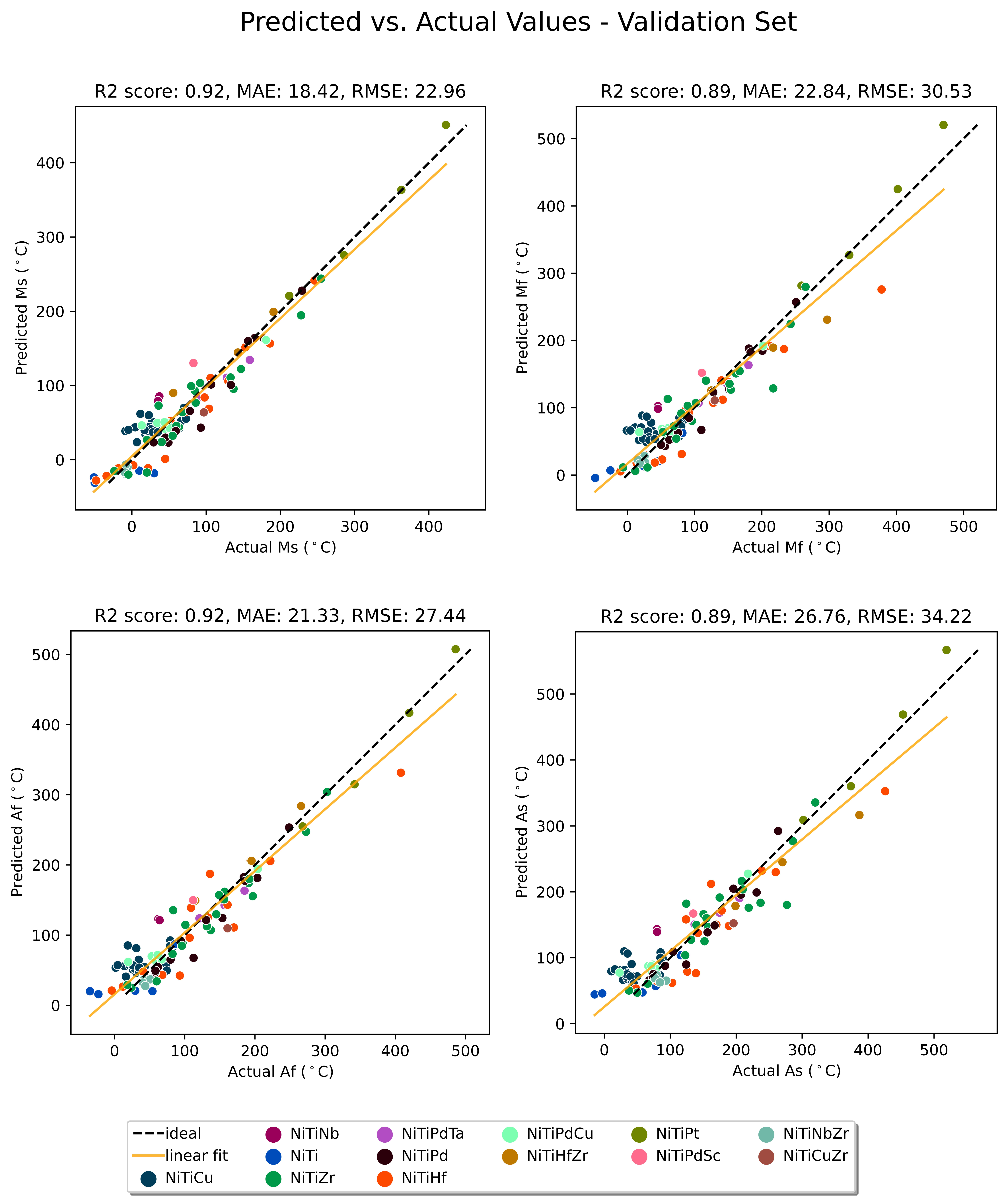}
  \caption{Predicted and actual values of Ms, Mf, As, and Af temperatures for the validation set. The figure was plotted using Matplotlib and seaborn. The visually distinct colors generated by iWantHue.}
  \label{fig:VS}
\end{figure}
We have further tested the model's performance on compositions unknown to the model to investigate how well this model can predict the transformation temperature of alloys, which the model has never seen during the previously described training/test procedure. This analysis is performed on the data set we referred to as the ``validation'' set with the distribution shown in Fig. \ref{fig:comp}. This data set contains 100 NiTi-based shape memory alloys, and the performance of the model on this data set is shown in Fig. \ref{fig:VS}. These data points are unique from the train/test data points either in the sense that they have a unique composition or processing parameters.  \par
\begin{figure*}[ht]
  \centering
  \includegraphics[width = \textwidth]{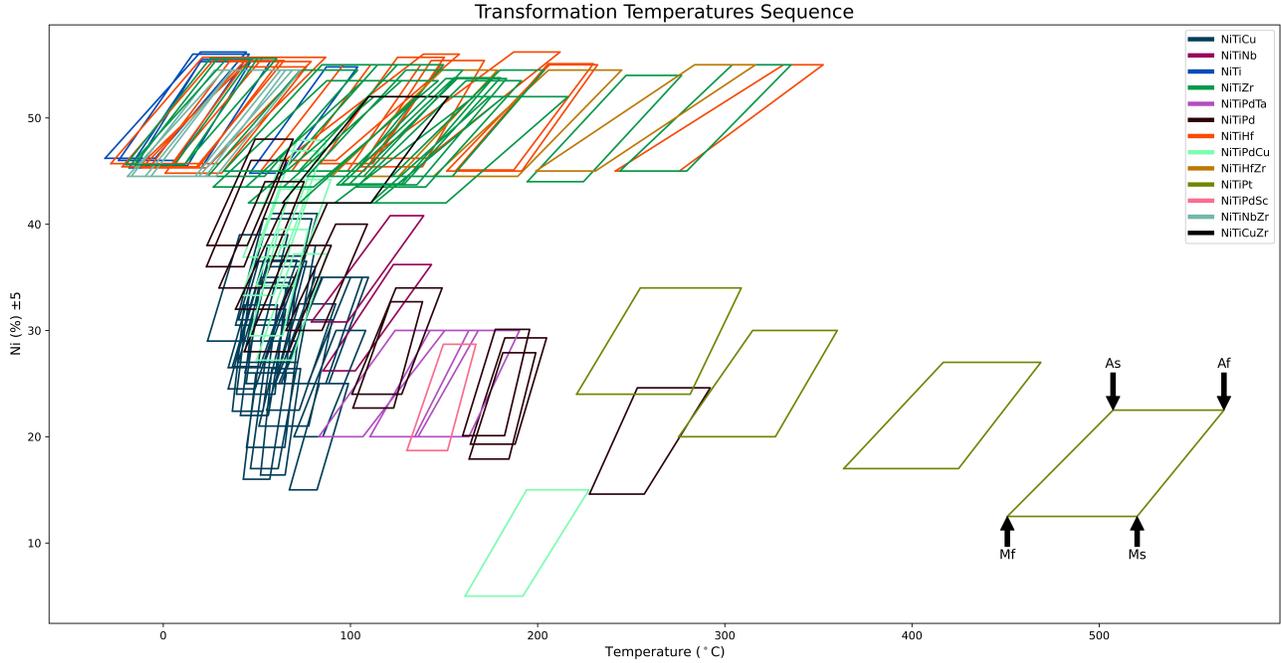}
  \caption{Transformation temperatures sequence for predicted values. Each parallelogram corresponds to a composition and each connection point corresponds to a transformation temperature. The figure was plotted using Matplotlib. The visually distinct colors generated by iWantHue.}
  \label{fig:sequence}
\end{figure*}
The comparison plot of the transformation temperatures for this validation set also shows the proximity of the predicted and true transformation temperatures. The $R^2$ score, MAE, and RMSE of the model for Ms temperature are 0.92, 18.42$^{\circ}C$, and 22.96$^{\circ}C$, for Mf temperature are 0.89, 22.84$^{\circ}C$, and 30.53$^{\circ}C$, for Af temperature are 0.92, 21.33$^{\circ}C$, and 27.44$^{\circ}C$, and for As temperature are 0.89, 26.76$^{\circ}C$, and 34.22$^{\circ}C$. \par 

It is worth mentioning that although the model was never trained with any NiTiNbZr alloy system, it could predict the targets for these alloy systems. Also, only 6 out of 26 alloy systems had more than 150 data points in the  training/test set, where the number of data points for the NiTiCu alloy system is lower than that of the validation set. Another interesting fact, as shown in Fig. \ref{fig:sequence}, is that the model could detect the sequence in transformation temperatures without defining any constraints. Accordingly, the model could understand and keep the following sequence: Ms $<$ Af, Mf $<$ As, As $<$ Af, and Mf $<$ Ms.

\begin{figure}[H]
  \centering
  \includegraphics[width = 0.5\textwidth]{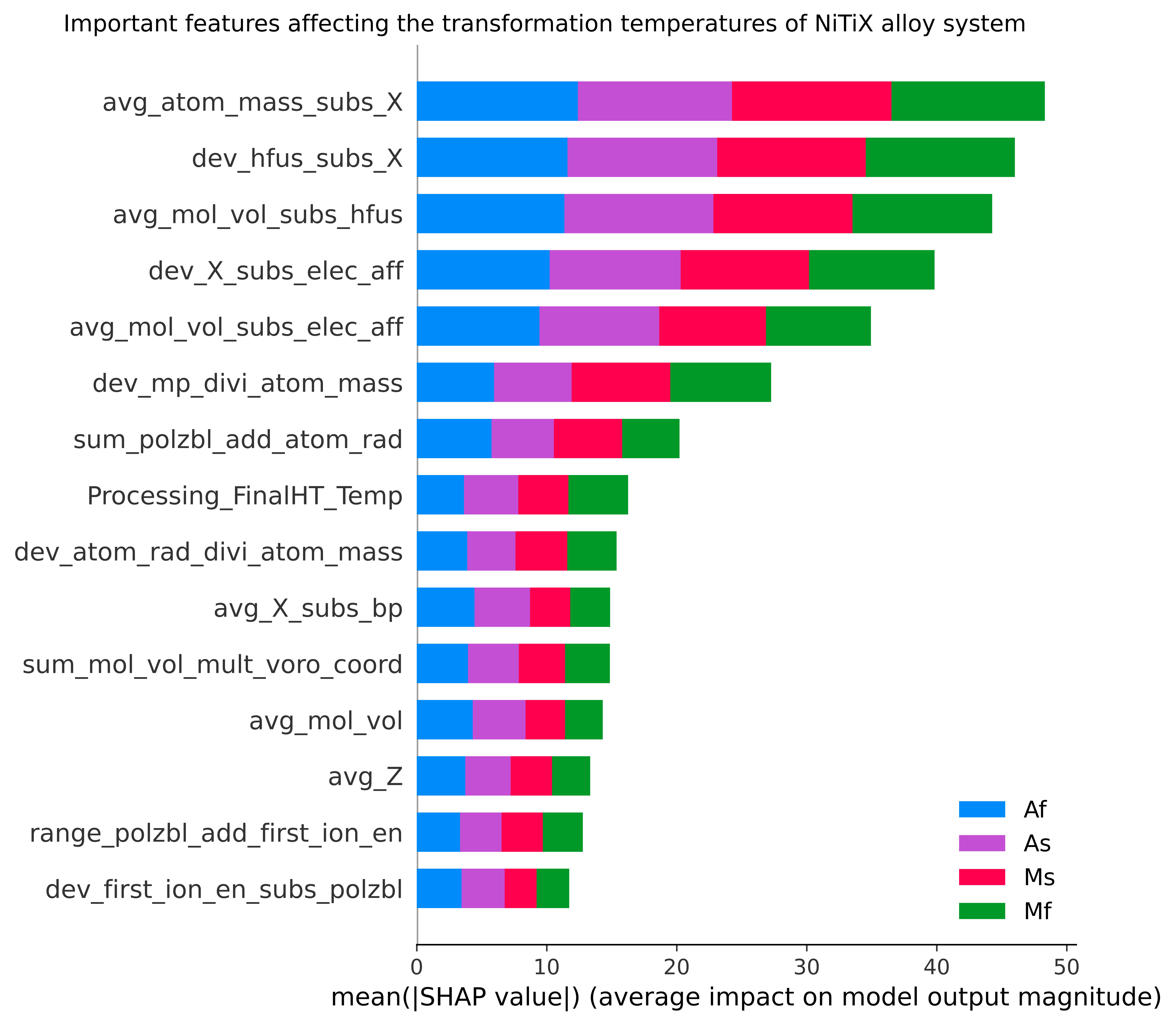}
  \caption{The most important features affect the transformation temperatures of the NiTiX alloy system. The figure was  plotted using the SHAP summary\_plot function.}
  \label{fig:if}
\end{figure}

Last but not least, the interpretation of the SHAP-based model was performed. The absolute Shapley values for each feature are averaged to create the SHAP feature importance plot, shown in Fig. \ref{fig:if}, which depicts the magnitude of feature attributions. This plot suggests that for the NiTiX alloy system, transformation temperatures are mainly a function of chemical and electronic properties such as atomic mass (atom\_mass), atomic radii (atom\_rad), molar volume (mol\_vol), heat of fusion (hfus), melting point (mp), boiling point (bp), electronegativity (X),  electron affinity (elec\_aff), and polarizability (polzbl). Moreover, the final heat treatment temperature (FinalHT\_Temp) as a processing parameter is among the top 10 most important features affecting this transformation temperature.

The uneven amount of SHAP values for targets (consider the width of the colorful area for each feature) suggests that not all features evenly affect transformation temperatures, which shows the complexity of the model in making predictions. \par
\begin{figure}[H]
  \centering
  \includegraphics[width = 0.5\textwidth]{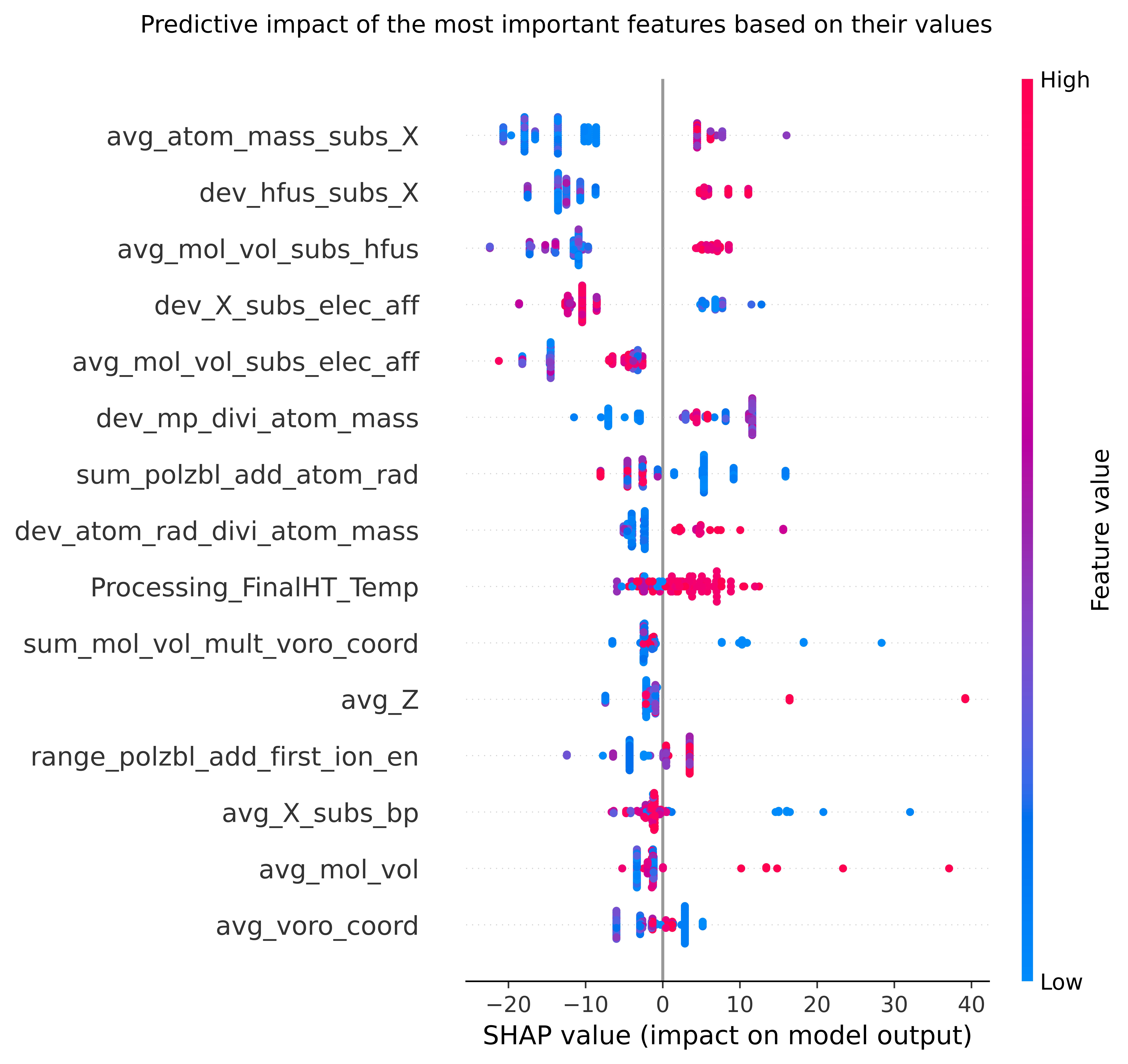}
  \caption{Effects of most important features' values on the Ms temperature of NiTiX alloys. The figure was plotted using the SHAP summary\_plot function.}
  \label{fig:ifMS}
  
\end{figure}

 
To better illustrate how changing values of a feature affect transformation temperature predictions, the summary plot for Ms temperature is depicted  in Fig. \ref{fig:ifMS} as an example. It shows that high values of average atomic mass and electronegativity difference (avg\_atom\_mass\_subs\_x), population standard deviation of heat of fusion and electronegativity difference
(dev\_hfus\_subs\_X), molar volume and heat of fusion difference (avg\_mol\_vol\_subs\_hfus),  population standard deviation of atomic radii and atomic mass quotient (dev\_atom\_rad\_divi\_atom\_mass) increase the $M_s$ temperature. On the contrary, low values of population standard deviation of electronegativity and electron affinity difference (dev\_X\_subs\_elec\_aff) and summation of polarizability and atomic radii (sum\_polzbl\_add\_atom\_rad) tend to increase this target. In the case of processing parameters, the high values of final heat treatment temperature can increase or decrease the martensitic temperature, based on its contrary effects on the microstructure and precipitates of each composition \cite{Frick2005ThermalAlloys}. Hence, it offers a handy tool for selecting alloying elements for designing SMAs with high transformation temperatures. 

When it comes to the computational costs, training the model for 2,394 data points with 3,087 features and 4 targets on a Windows laptop equipped with Intel\textsuperscript{\textregistered} Core\textsuperscript{\texttrademark} i7-11800H processor and 16 GB Random Access Memory (RAM) took 3 minutes and 21 seconds. The model proves to be fast, accurate, and simple - it has competitive performance and can be an excellent choice for time-sensitive studies.

\section{Conclusions}
Employing advanced ML techniques and domain knowledge, we were able to build a model to predict transformation temperatures of SMAs, provide insight into the important features contributing to each prediction, and shed light on the physics of thermodynamic transformation of SMAs. The results of the present work are as follows:
\begin{itemize}
\item Proper feature engineering with all processing parameters, composition-based features, and alloy system categorical features is the key to building a robust model that is able to predict transformation temperatures of a vast range of SMAs with a relatively low amount of data. In our study, the model could predict the transformation temperatures of 100 SMAs in the validation set after training by 2,394 data points for 137 alloys in the train/test set.
  
  \item The results shows that CatBoost is a promising ML surrogate model for the thermodynamic prediction of complex alloy systems. Considering its tree-based nature, it can adequately handle diverse data sets with a potentially high number of features, which is the case in most materials science studies. It also performs well with a relatively limited number of training data points. This tree-based boosting algorithm also provides a natural path to model  interpretability. 
  \item The dependency of transformation temperatures on processing parameters, chemical, and electronic properties was shown and discussed.

\end{itemize}
\section*{Data availability}
Data used in this study is part of an ongoing research project and cannot be shared publicly at this time.
\section*{Acknowledgements}
The present work is supported by the DMREF NSF Grant No. 2119103 and 1905325. WT acknowledges the support of NSF through Grant No. NSF-DGE-1545403. GV acknowledges the support from QNRF through Grant No. NPRP11S-1203-170056. High-Performance Research Computing at Texas A\&M University provided computing resources for ML calculations.

\bibliographystyle{IEEEtran}
\bibliography{references.bib, manualreferences.bib}{}
\end{multicols}
\end{document}